\documentclass{PoS}
\title{Large parton densities and high$-p_T$ physics in heavy-ion collisions}
\ShortTitle{Large parton densities and high$-p_T$ physics in heavy-ion collisions}
\author{Cyrille Marquet\\
Institut de Physique Th\'eorique, CEA/Saclay, 91191 Gif-sur-Yvette cedex, France\\
Department of Physics, Columbia University, New York, NY 10027, USA
\\E-mail:\email{cyrille@phys.columbia.edu}}

\abstract{
I discuss the role played by large parton densities in the QCD description of
high$-p_T$ observables in relativistic heavy-ion collisions. In pA collisions, high$-p_T$ particles probe large parton densities in the nucleus and provide tests of the Color Glass Condensate (CGC) picture of the small$-x$ part of the nuclear wave function. In AA collisions, high$-p_T$ particles carry information on both the Quark-Gluon-Plasma phase, and the Glasma phase created at early times by the collision of two CGCs.}

\FullConference{4th international workshop High-pT physics at LHC 09\\
February 4-7, 2009\\Prague, Czech Republic}

\begin{document}

\section{Large parton densities in the nuclear wave function}

\begin{figure}[t]
\begin{center}
\includegraphics[width=5cm,angle=-90]{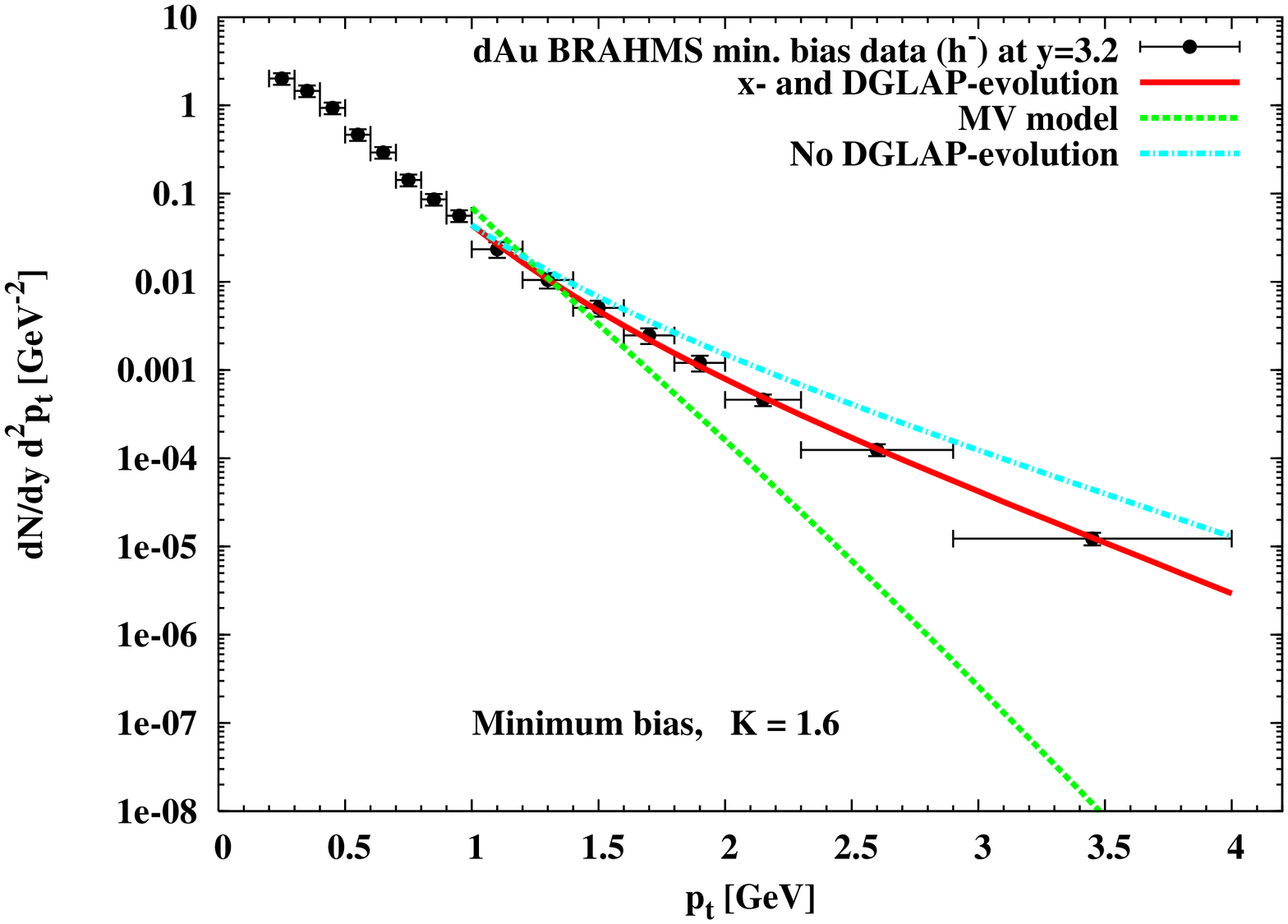}
\hspace{0.5cm}
\includegraphics[width=5cm,angle=-90]{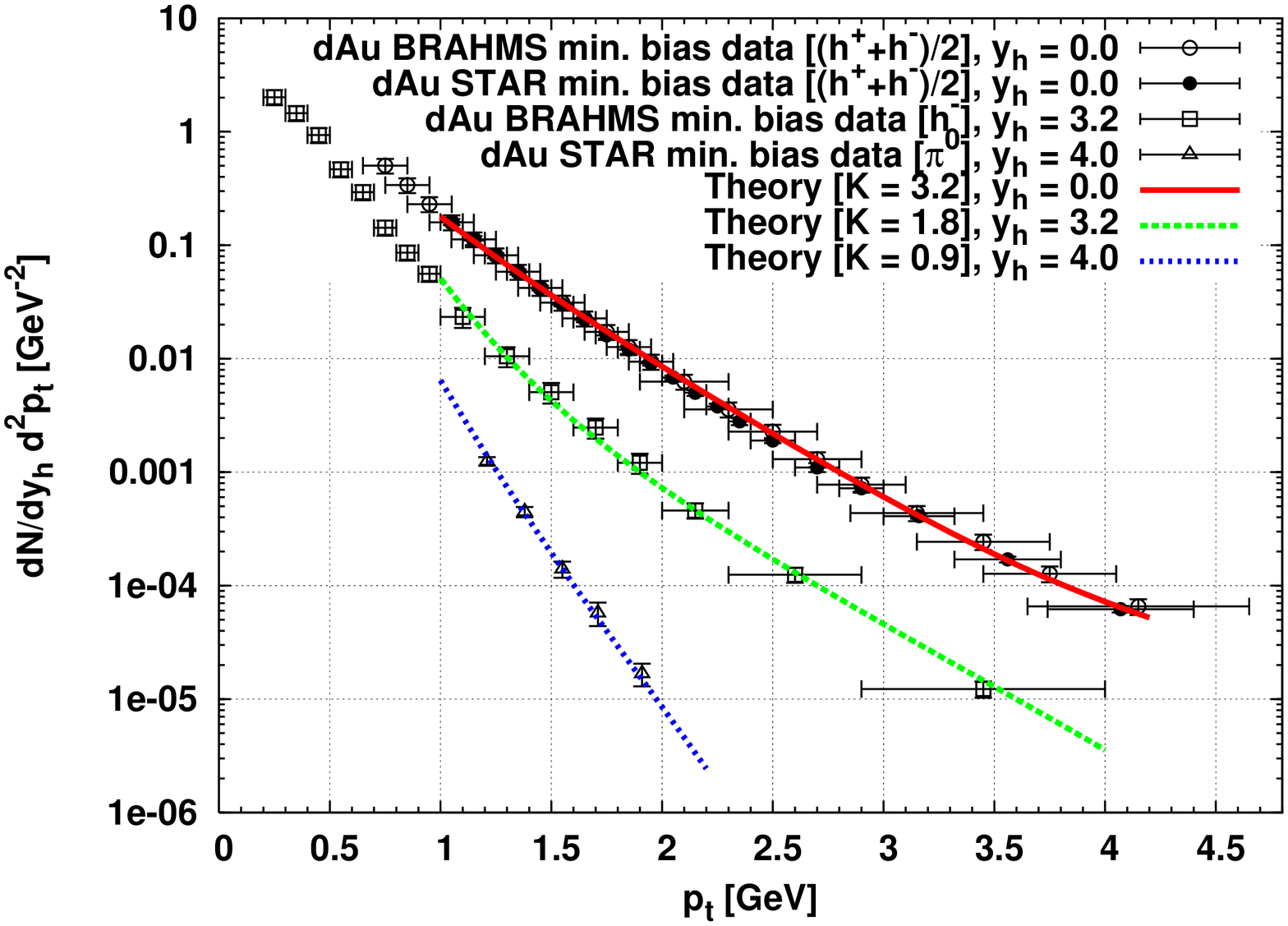}
\caption{Forward particle production in d+Au collisions at RHIC. The left plot shows the importance of including both the large-$x$ DGLAP evolution of the dilute deuteron and the
small-$x$ CGC evolution of the dense nucleus. The right plots shows the excellent description of the spectra shapes, and the K factors needed to obtain the normalization.}
\end{center}
\end{figure}

When probing small distances inside a hadron or nucleus with a hard process, one resolves their partonic constituents. Increasing the energy of the scattering process at a fixed momentum transfer allows to probe lower-energy partons, with smaller energy fraction $x.$ As the parton densities in the hadronic/nuclear wave function grow with decreasing $x,$ they eventually become so large that a non-linear (yet weakly-coupled) regime is reached, called saturation, where partons do not interact with the probe independently anymore, but rather behave coherently. 

The Color Glass Condensate (CGC) is an effective theory of QCD \cite{cgcrev} which aims at describing this part of the wave function. Rather than using a standard Fock-state decomposition, it is more efficient to describe it with collective degrees of freedom, more adapted to account for the collective behavior of the small-$x$ gluons. The CGC approach uses classical color fields: 
\begin{equation}
|h\rangle=|qqq\rangle+|qqqg\rangle+\dots+|qqqg\dots ggg\rangle+\dots\quad
\Rightarrow\quad|h\rangle=\int D\rho\ \Phi_{x_A}[\rho]\ |\rho\rangle
\label{cgc}\ .\end{equation}
The long-lived, large-$x$ partons are represented by a strong color source
$\rho\!\sim\!1/g_S$ which is static during the lifetime of the short-lived small-$x$ gluons, whose dynamics is described by the color field $A^\mu\!\sim\!1/g_S.$ The arbitrary separation between the field and the source is denoted $x_A.$

When probing the CGC with a dilute object carrying a weak color charge, the color field $A^\mu$ is directly obtained from $\rho$ via classical Yang-Mills equations:
\begin{equation}
[D_\mu,F^{\mu\nu}]=\delta^{-\nu}\rho\ ,
\end{equation}
and it can be used to characterize the CGC wave function $\Phi_{x_A}[A^-]$
(in the $A^+\!=\!0$ gauge). This wave function is a fundamental object of this picture, it is mainly a non-perturbative quantity, but the $x_A$ evolution can be computed perturbatively. Requiring that observables are independent of the choice of $x_A,$ a functional renormalization group equation can be derived. In the leading-logarithmic approximation which resums powers of $\alpha_S\ln(1/x_A),$ the JIMWLK equation describes the evolution of $|\Phi_{x_A}|^2$ with $x_A.$

The information contained in the wave function, on gluon number and gluon correlations, can be expressed in terms of n-point correlators, probed in scattering processes. These correlators consist of Wilson lines averaged with the CGC wave function, and resum powers of $g_S A^-$ (therefore both multiple scatterings and non-linear QCD evolution are taken into account). For instance in the case of single inclusive gluon production in pA collisions, the CGC is described by its (all-twist) unintegrated gluon distribution, obtained from the 2-point function \cite{mygprod}. More exclusive observables involve more complicated correlators. 

\begin{figure}[t]
\begin{center}
\includegraphics[width=6.8cm]{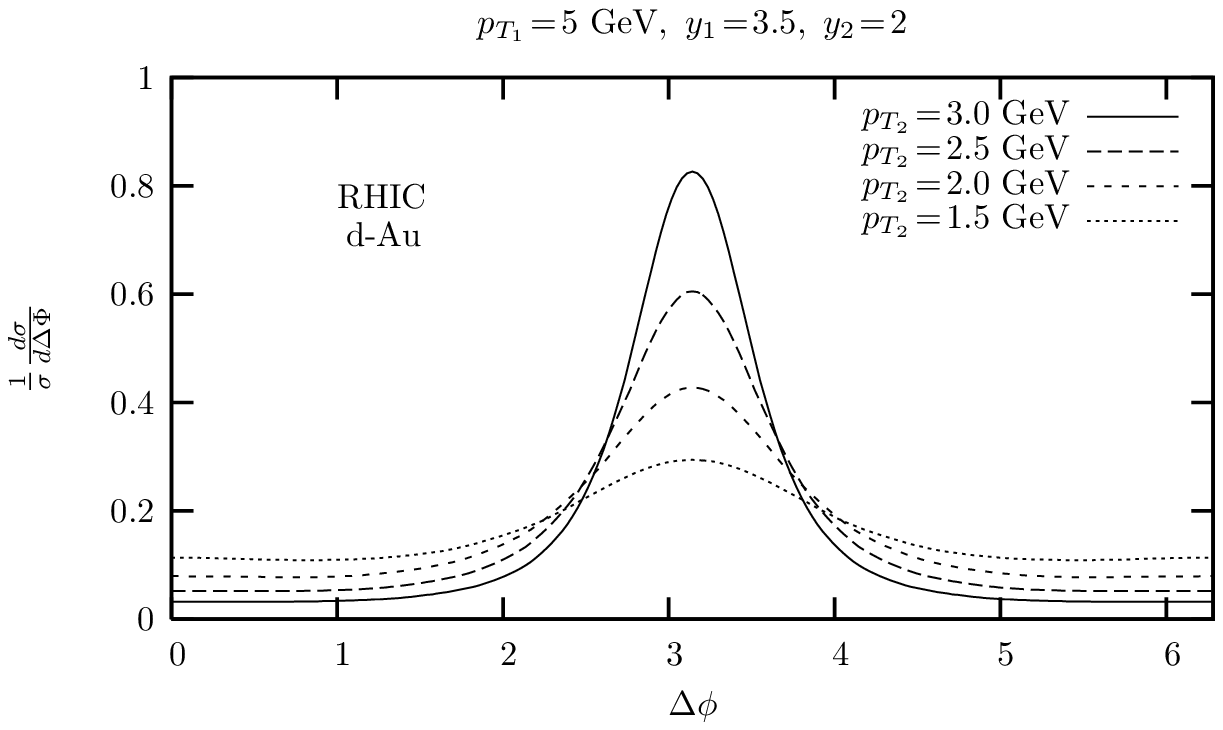}
\hspace{0.5cm}
\includegraphics[width=6.8cm]{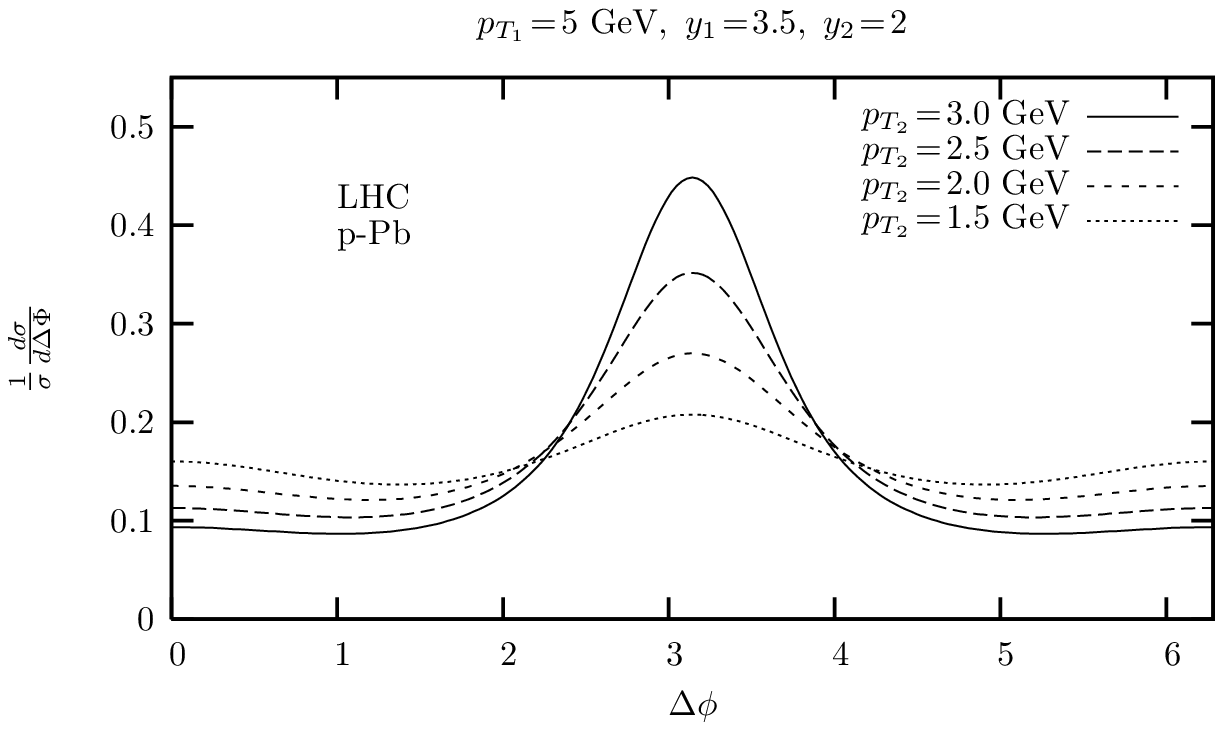}
\caption{Two-particle production at forward rapidities in pA collisions. The $\Delta\phi$ spectrum is displayed at RHIC (left) and LHC (right) energies. When decreasing $p_{T_2}$ at fixed $y_2,$ the correlation in azimuthal angle is suppressed. At the LHC, smaller values of $x_A$ are probed, and the azimuthal angle correlation is more suppressed as indicated by the vertical axis; the peak is also less pronounced.}
\end{center}
\end{figure}

Forward particle production in pA collisions allows to investigate the non linear QCD dynamics of high-energy nuclei with a probe well understood in QCD. Indeed, while such processes are probing small-momentum partons in the nuclear wave function, only high-momentum partons of the proton contribute to the scattering ($\sqrt{s} x_p\!=\!k e^y$ and $\sqrt{s} x_A\!=\!k e^{-y}$ with $k$ and $y$ denoting transverse momentum and rapidity), and that involves standard parton distribution functions. In two-particle production, contrary to single particle production, the CGC cannot be described only by its unintegrated gluon distribution, the so-called $k_T$-factorization framework is not applicable.

It was not obvious that the CGC picture (\ref{cgc}), which requires small values of $x_A,$ would be relevant at present energies. One of the most acclaimed successes came in the context of d+Au collisions at RHIC: the prediction that the yield of high-$p_T$ particles at forward rapidities in d+Au collisions is suppressed compared to A pp collisions, and should decrease when increasing the rapidity, was confirmed
\cite{jyrev}. In Fig.1 the $dAu\!\to\!hX$ $p_T$ spectra computed in the CGC approach \cite{dhj} is compared to RHIC data, and the description of the slope is impressive. The need of K factors to describe the normalization could be expected since this is a leading-order based calculation. Improving the calculation with the next-leading evolution has yet to be done.

The focus should now shift towards more exclusive observables like two-particle production $pA\!\to\!h_1h_2X.$ In particular the correlations in azimuthal angle between the produced hadrons should be suppressed compared to pp collisions. Predictions for the process $pA\to h_1h_2X$ are shown in Fig.2, for RHIC and the LHC \cite{mytpc}. $k_1,$ $k_2$ and $y_1,$ $y_2$ are the transverse momenta and rapidities of the final state hadrons, and the azimuthal angle spectra are displayed. It is obtained that the perturbative back-to-back peak of the azimuthal angle distribution is reduced by initial state saturation effects. As the momenta decrease, the angular distribution broadens.

\section{Particle production in the Glasma}

\begin{figure}[t]
\begin{center}
\includegraphics[width=6.5cm]{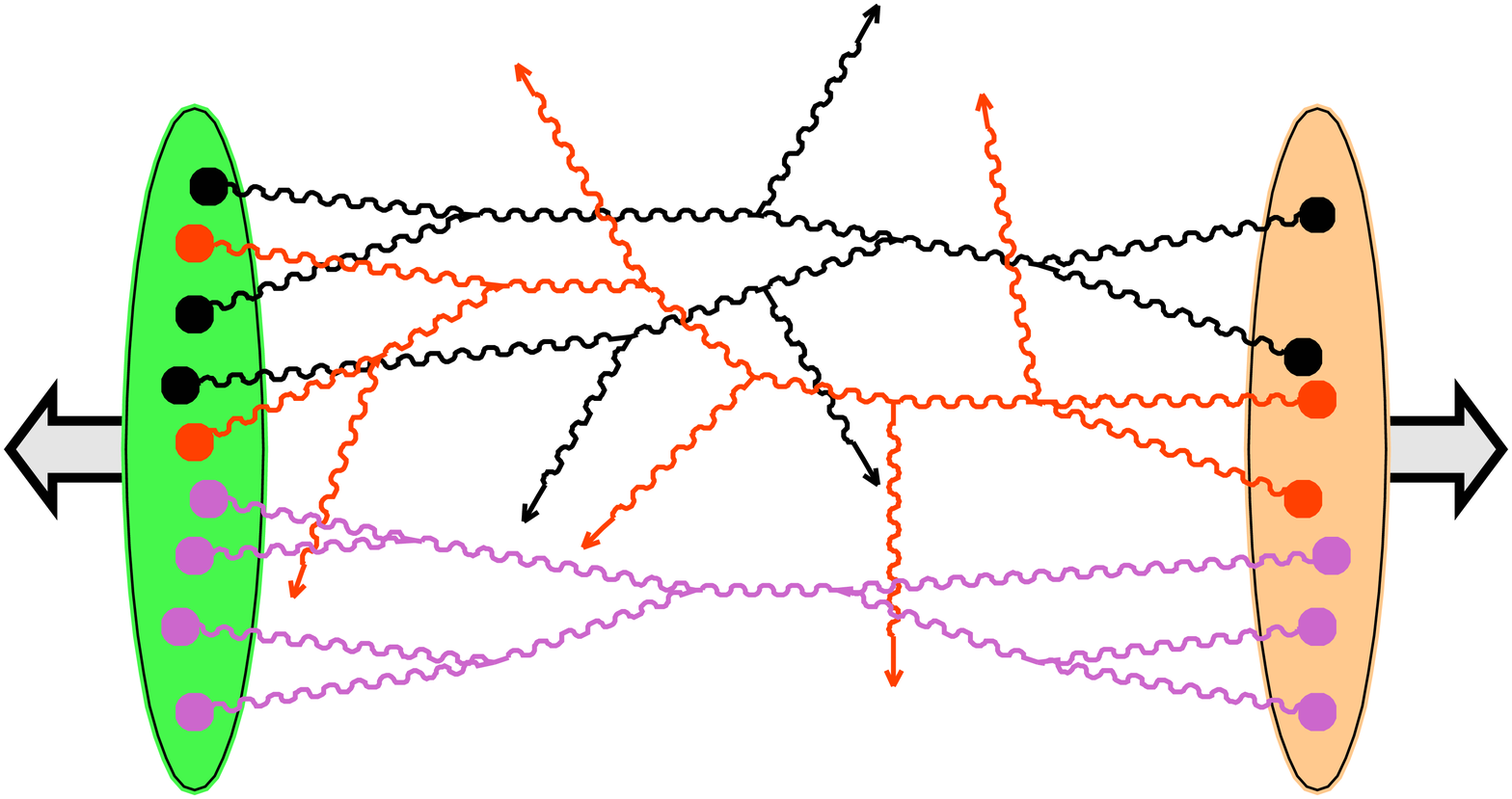}
\hspace{0.5cm}
\includegraphics[width=7.5cm]{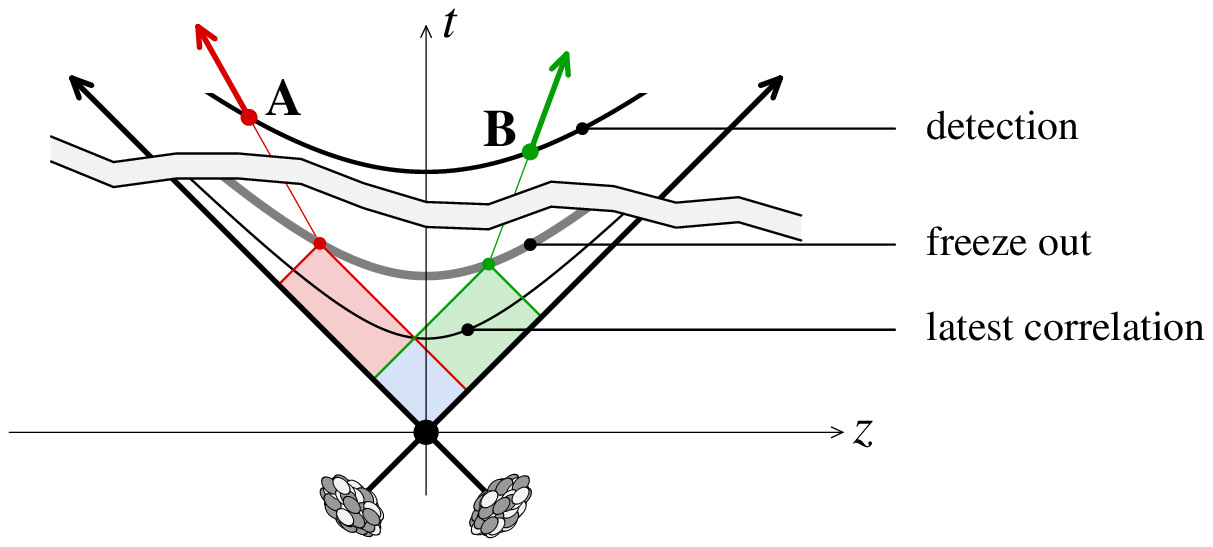}
\caption{Left: typical leading-order diagram for particle production in the Glasma, multiple partonic interactions are crucial when low values of $x$ are being probed in the nuclear wave functions. Right: the space-time location of events that may correlate two particles is the intersection of their past light-cones. Correlations between particles widely separated in rapidity are due to early-time dynamics.}
\end{center}
\end{figure}

The Glasma is the result of the collision of two CGCs, and it provides a weak-coupling description of the early stages after a high-energy heavy-ion collision. Each nuclear wave function is characterized by a strong color charge, and the field describing the dynamics of the small-x gluons is the solution of
\begin{equation}
[D_\mu,F^{\mu\nu}]=\delta^{+\nu}\rho_1+\delta^{-\nu}\rho_2\ .
\end{equation}
The field after the collision is non-trivial \cite{glasma}: it has a strong component ($A^\mu\sim1/g_s$), a component which is particle like ($A^\mu\sim1$), and components of any strength in between. To understand how this pre-equilibrium system thermalizes, one needs to understand how the Glasma field decays into particles. Right after the collision, the strong field component contains all modes.
Then, as the field decays, modes with $p_T>1/\tau$ are not part of the strong component anymore, and for those a particle description becomes more appropriate. After a time of order $1/Q_s,$ this picture breaks down, and it has been a formidable challenge to determine weather a fast thermalization can be achieved within this framework.

A problem which can be more easily addressed is particle production. The difficult task is to express the cross-section in terms of the Glasma field, taking into account multiple partonic interactions, as pictured in Fig.3 (left). Because of the flux-tube structure of its color field $A^\mu,$ the Glasma is a natural candidate to explain the ridge-shaped two-particle correlations observed at RHIC, as well as three-particle correlations \cite{ridge}. The ridge is collimated in azimuthal angle because of the radial flow which happens at a later stage, but since the ridge is several units long in rapidity, it is due to early time dynamics: this is explained in Fig.3 (right) which shows the space-time picture of the collision. In the forward light-cone, lines of constant proper time $\tau=\sqrt{x^+x^-}$ are hyperbolae and lines of constant rapidity $\eta=\frac12\log(x^+/x^-)$ are straight lines from the origin. For two final-state particles separated by the rapidity $\Delta\eta,$ causality imposes that they can be correlated only by events which happened at
\begin{equation}
\tau<\tau_{f.o.}\ e^{-\Delta\eta/2}\ ,
\end{equation}
where the freeze-out time $\tau_{f.o.}$ denote the time of last interaction. While the features of the ridge are qualitatively explained by the Glasma, a quantitative description is needed.

\section{Energy loss of high-$p_T$ partons in the QCD plasma}

Hard probes are believed to be understood well enough to provide clean measurements of the properties of the QGP formed in heavy-ion collisions. A large amount of work has been devoted to understand what happens to a quark (of high energy $E,$ mass $M$ and Lorentz factor  $\gamma=E/M$) as it propagates through a thermalized plasma
\cite{jqrev}. Multiple scatterings are a main ingredient of the perturbative QCD (pQCD) description of how a quark losses energy, until it thermalizes or exits the medium (see Fig.4).

At lowest order with respect to $\alpha_s,$ quantum fluctuations in a quark wave function consist of a single
gluon, whose energy we denote $\omega$ and transverse momentum $k_\perp.$ The virtuality of that
fluctuation is measured by the coherence time, or lifetime, of the gluon $t_c=\omega/k_\perp^2.$
Short-lived fluctuations are highly virtual while longer-lived fluctuations are more easily put on shell when
they interact. The probability of the fluctuation is $\alpha_sN_c,$ up to a kinematic factor which for heavy
quarks suppresses fluctuations with $\omega>\gamma k_\perp.$ This means that when gluons are put
on-shell, they are not radiated in a forward cone around a heavy quark. This suppression of the available
phase space for radiation, the {\it dead-cone} effect, implies less energy loss for heavier quarks.

\begin{figure}[t]
\begin{center}
\includegraphics[width=4.5cm]{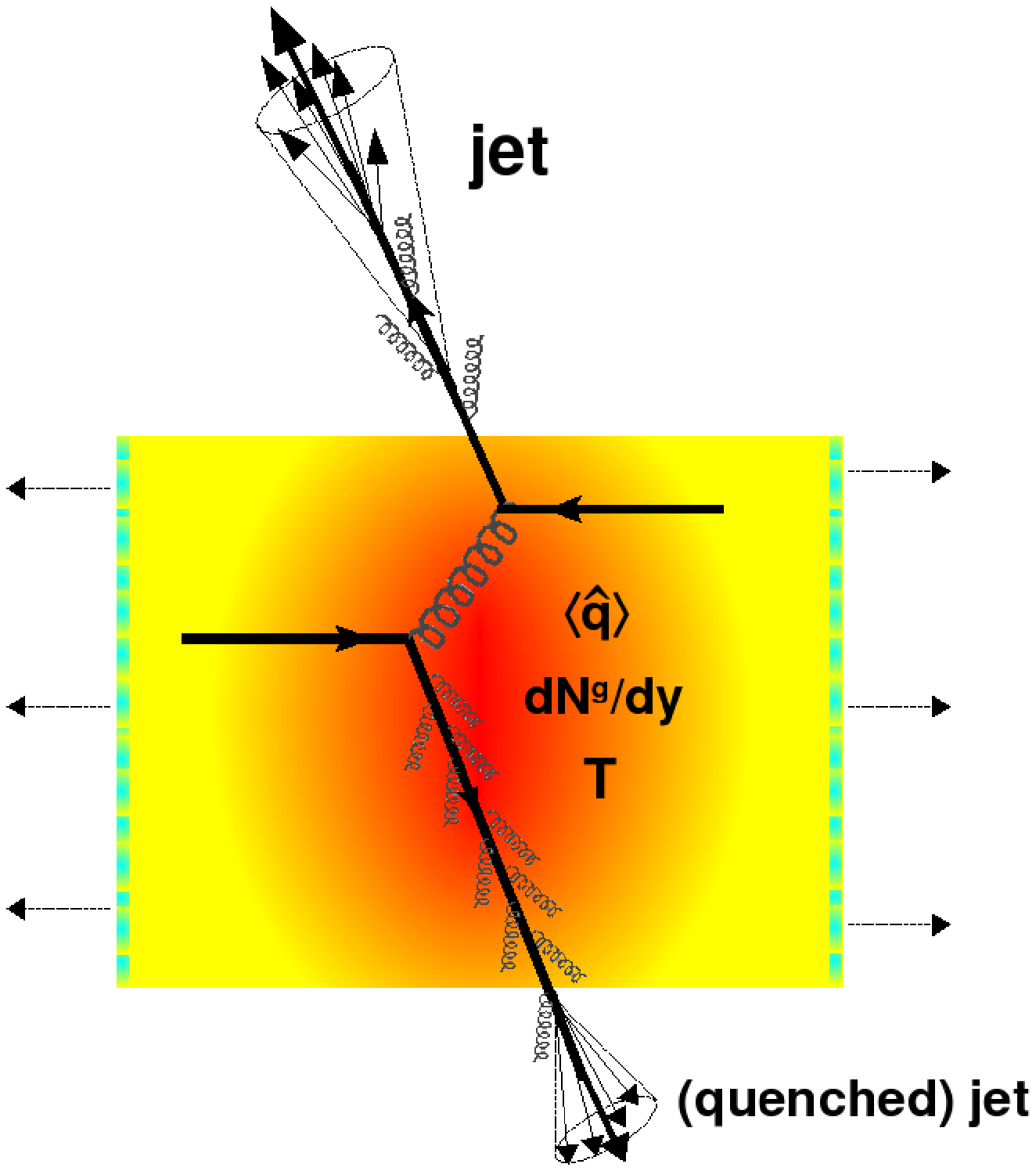}
\hspace{1cm}
\includegraphics[width=6.5cm]{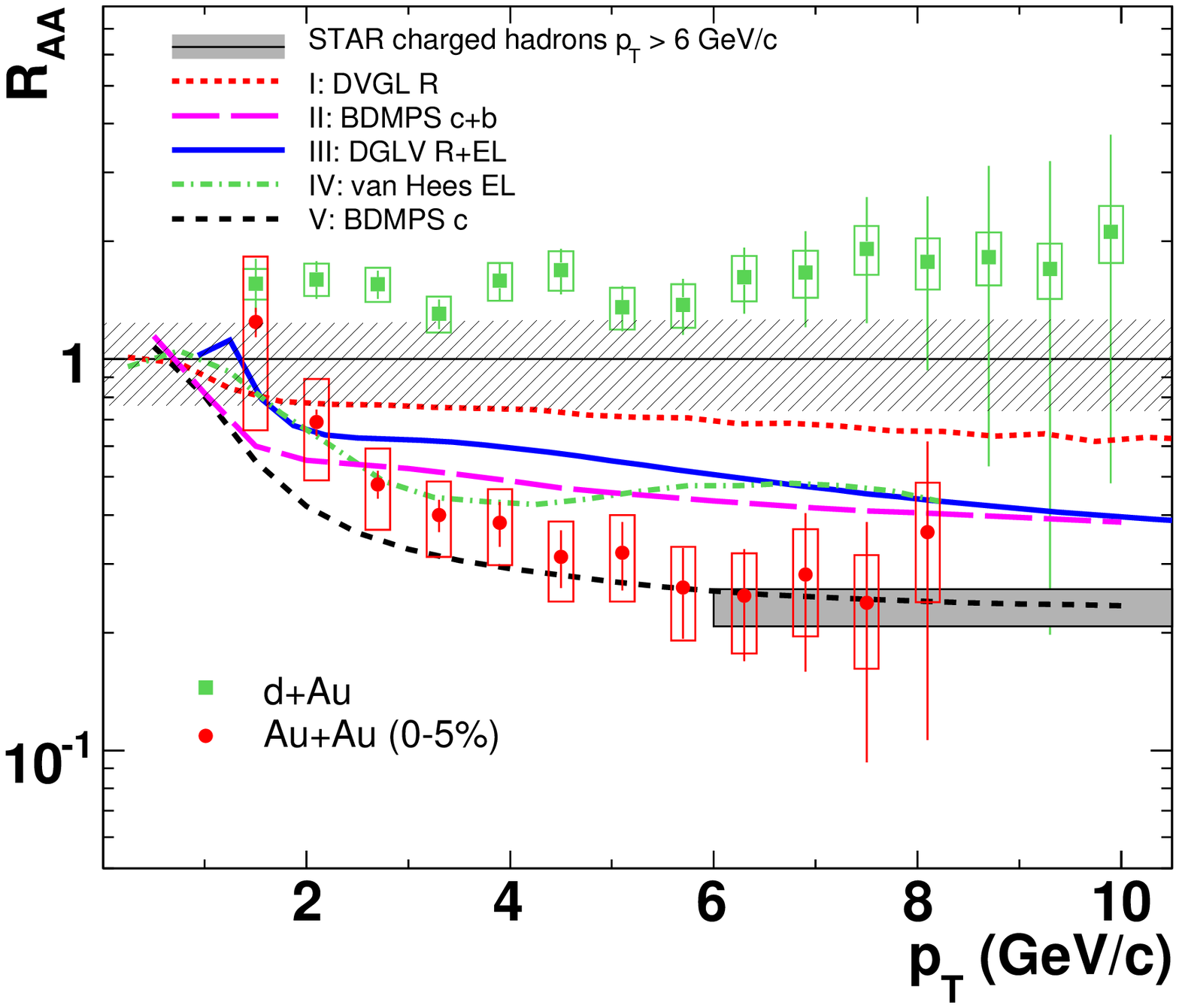}
\caption{Left: production of high-energy partons in a hard process, which then lose energy propagating through the plasma. Some quantum fluctuations in their wave function are put on shell while interacting with the medium and become emitted radiation.
Right: the resulting particle production in AA collisions is suppressed ($R_{AA}<1$) compared to independent nucleon-nucleon collisions. The suppression is large for light hadrons, and similar for heavy mesons (those data are displayed in the figure), which is difficult to accommodate in a weakly-coupled QCD description.}
\end{center}
\end{figure}

In pQCD, medium-induced gluon radiation is due to multiple scatterings of the virtual gluons.
If, while undergoing multiple scattering, the virtual gluons pick up enough transverse momentum to be put on shell,
they become emitted radiation. The accumulated transverse momentum squared picked up by a gluon of coherence
time $t_c$ is
\begin{equation}
p_\perp^2=\mu^2 \frac{t_c}{l}\equiv\hat{q}\ t_c
\end{equation}
where $\mu^2$ is the average transverse momentum squared picked up in each scattering, and
$l$ is the mean free path. These medium properties are involved through the ratio
$\hat{q}=\mu^2/l.$

Since only the fluctuations which pick up enough transverse momentum are freed ($k_\perp<p_\perp$),
the limiting value can be obtained by equating $k_\perp^2$ with
$p_\perp^2=\hat{q}\omega/k_\perp^2:$
\begin{equation}
k_\perp<(\hat{q}\omega)^{1/4}\equiv Q_s(\omega)\ .
\end{equation}
The picture is that highly virtual fluctuations with $k_\perp>Q_s$ do not have time to pick up enough $p_\perp$ to be freed, while the longer-lived ones with $k_\perp<Q_s$ do. That transverse momentum $Q_s$ which controls which gluons are freed and which are not is called the saturation scale. With heavy quarks, one sees that due to the dead cone effect, the maximum energy a radiated gluon can have is $\omega=\gamma k_\perp=\gamma Q_s$ (and its coherence time is $t_c=\gamma/Q_s)$. This allows to
estimate the heavy-quark energy loss:
\begin{equation}
-\frac{dE}{dt}\propto\alpha_sN_c\frac{\gamma Q_s}{\gamma/Q_s}=\alpha_s N_c Q_s^2\ .
\label{eloss}\end{equation}
The saturation momentum in this formula is the one that corresponds to the fluctuation which dominates the energy loss: $Q_s=(\hat{q}\gamma)^{1/3}.$

For a plasma of extend $L\!<\!t_c=\gamma^{2/3}/\hat{q}^{1/3},$ formula (\ref{eloss}) still holds but with $Q_s^2=\hat{q}L.$ These are the basic ingredients of more involved phenomenological calculations, but after comparisons with data, it has remained unclear if this perturbative approach can describe the suppression of high$-p_\perp$ particles. For instance, at RHIC temperatures, the value $\hat{q}\sim 1-3\ \mbox{GeV}^2/\mbox{fm}$ is more natural than the $5-10\ \mbox{GeV}^2/\mbox{fm}$ needed to describe the data on light hadron production. If one accepts to adjust $\hat{q}$ to this large value, then the $D$ and $B$ mesons are naturally predicted to be less suppressed than light hadrons, which is not the case (see Fig.4).

While the present pQCD calculations should still be improved, and may be shown to work in the future, this motivated to think about strongly-coupled plasmas. The tools to address the strong-coupling dynamics in QCD are quite limited, however for the $N=4$ Super-Yang-Mills (SYM) theory, the AdS/CFT correspondence is a powerful approach used in many studies. The findings for the strongly-coupled SYM plasma may provide insight for gauge theories in general, and some aspects may even be universal. One interesting result is that the total energy loss of hard probes goes as $\Delta E\propto L^3$ at strong coupling \cite{us}, instead of the $L^2$ law at weak coupling.

\end{document}